\documentclass[amsmath,amssymb,twocolumn]{revtex4}

\usepackage{graphicx}

\begin{document}

\title{Raman spectroscopy on cubic and hexagonal SrMnO$_3$}
\author{A. Sacchetti\footnote{corresponding author. Email:
sacchetti@phys.ethz.ch, Tel.: +41 44 6332241, Fax: +41 44
6331072}, M. Baldini, P. Postorino} \affiliation{INFM and
Dipartimento di Fisica, Universit\`a ``La Sapienza'', P.le A. Moro
4, I-00187 Roma, Italy.}
\author{C. Martin, A. Maignan}
\affiliation{Laboratoire CRISMAT-UMR, 6508 ENSI CAEN, 6, Marechal
Juin, 14050 Caen, France.}

\begin{abstract}
We report on the first optical characterization of both cubic and
hexagonal SrMnO$_3$. Room-temperature Raman spectra collected by
means of a micro-Raman spectrometer are shown. The spectrum of the
cubic compound is characterized by weak and broad bands in
agreement with group-theory which predicts no Raman-active phonons
for this compound. On the other hand, the spectrum of the
hexagonal compound shows six narrow peaks ascribed to one-phonon
processes. A complete polarization analysis of the spectra
collected from a single crystallite allows us to completely assign
the symmetries of the six observed peaks. Atomic displacements for
each phonon peaks are also proposed.
\end{abstract}

\maketitle

\section{INTRODUCTION}

In the last years, a remarkable experimental and theoretical
effort was devoted to the study of colossal magneto-resistance
manganites $A_{1-x}A'_x$MnO$_3$ ($A$ is a trivalent rare-earth and
$A'$ is a divalent alkaline earth). The peculiar properties of
these systems have attracted much attention owing to their
possible technological applications in the field of magnetic
devices and, on the side of basic physics, to the possibility of
disentangling the delicate interplay among the different
microscopic interactions which govern their often uncommon
behavior \cite{generali,generali2}. The $T-x$ (temperature-doping)
phase diagram of manganites is typically very rich and presents a
large number of structural, magnetic, and conducting phases
\cite{dagotto}. Experimental and theoretical studies on doped
compounds ($0<x<1$) have contributed to build up a rather general
framework, where a key role is played by the oxygen-mediated
magnetic interactions among the Mn ions and the electron-phonon
coupling triggered by a Jahn-Teller distortion of the MnO$_6$
octahedra. In particular, the strength of the charge-lattice
coupling is directly related to the average Mn valence state (i.e.
to $x$) and to the spatial arrangements of the MnO$_{6}$
sub-lattice. Much less attention was devoted to the $x=1$ end
compounds and there are very few studies, mainly appeared in early
papers, reporting on SrMnO$_{3}$ (SMO), despite its importance.
SMO is indeed the parent compound of one of the most studied
manganite family (La$_{1-x}$Sr$_{x}$MnO$_{3}$) and of electron
doped manganites series such as Sr$_{1-x}$Ce$_{x}$MnO$_{3}$, which
are nowadays of utmost interest. Moreover, the possibility of
switching the lattice symmetry of SMO from cubic to hexagonal
depending on the growing parameters could give unique
opportunities of studying the role of the spatial arrangements of
the MnO$_{6}$ network. Since optical spectroscopies have been
largely and successfully employed to study manganites
\cite{raman}, in the present paper we report the first Raman
characterization of cubic (cSMO) and hexagonal (hSMO) SrMnO$_{3}$
and a complete assignment of the observed Raman active modes.

\section{EXPERIMENTAL}

The SMO polycrystalline samples were synthesized with standard
solid-state reaction from stoichiometric mixtures of SrCO$_3$ and
MnO$_2$ at $1500^\circ$C \cite{Hervieu,Chmaissem}. The sample
synthesized in air is characterized by an hexagonal structure. The
oxygen-deficient cubic compound was obtained in Ar-flow. Then, the
cubic compound was annealed for $24$ h at $600^\circ$C, in a $100$
bar O$_2$ atmosphere in order to reach the O$_3$ stoichiometry. A
detailed characterization of both sample was performed by means of
X-ray diffraction and magnetization measurements. The hexagonal
and cubic compounds belong to the $P6_{3}/mmc$ and $Pm3m$ space
groups respectively, both samples showing a low-temperature
antiferromagnetic  phase with a Ne\'{e}l temperature of $278$ K
(hexagonal) and $260$ K (cubic) \cite{Battle,takeda}.

Raman-scattering measurements were performed in the backscattering
geometry using a confocal Raman microspectrometer (Labram
Infinity, Jobin Yvon). The spectrometer was equipped with an He-Ne
laser source (we used the $632.81$ nm line with a power of about
$15$ mW on the sample surface) and a cooled charge coupled device
(CCD) detector to collect the scattered light dispersed by a
$1800$ lines/mm grating. The notch filter, used to reject the
elastically scattered light, does not allow reliable measurements
below $200$ cm$^{-1}$. The microscope equipped with a $20 \times$
objective enables for a spatial resolution gain of a few microns
on the sample surface. Room temperature Raman spectra of SMO
(hexagonal and cubic), with a resolution of about $3$ cm$^{-1}$,
were collected over the $200$-$1100$ cm$^{-1}$ spectral range.

Sample's homogeneity was checked by comparing the spectra
collected by focusing the laser beam on different spots of the
samples surface. The Raman spectra of cSMO have shown a quite good
reproducibility whereas remarkable fluctuations of the band
intensities were observed from point to point in the hSMO. Bearing
in mind that the incident beam has a definite linear polarization,
the presence of crystallites randomly oriented on the micron scale
can account for the observed fluctuations. Optical and
spectroscopic analysis of very small fragments of our hSMO sample
confirmed this hypothesis and allowed us to select some small
single crystals to be analyzed. A polarization analysis of the
Raman signal was thus carried out on a very small single crystal
of the hSMO (about $100 \times 20$ $\mu$m$^{2}$) using a rotating
platform with fine tilting movements to carefully align the sample
surface and a polarization analyzer placed just in front of the
monochromator entrance slit.

\section{RESULTS AND DISCUSSION}

\begin{figure}[!t]
\includegraphics[width=8cm]{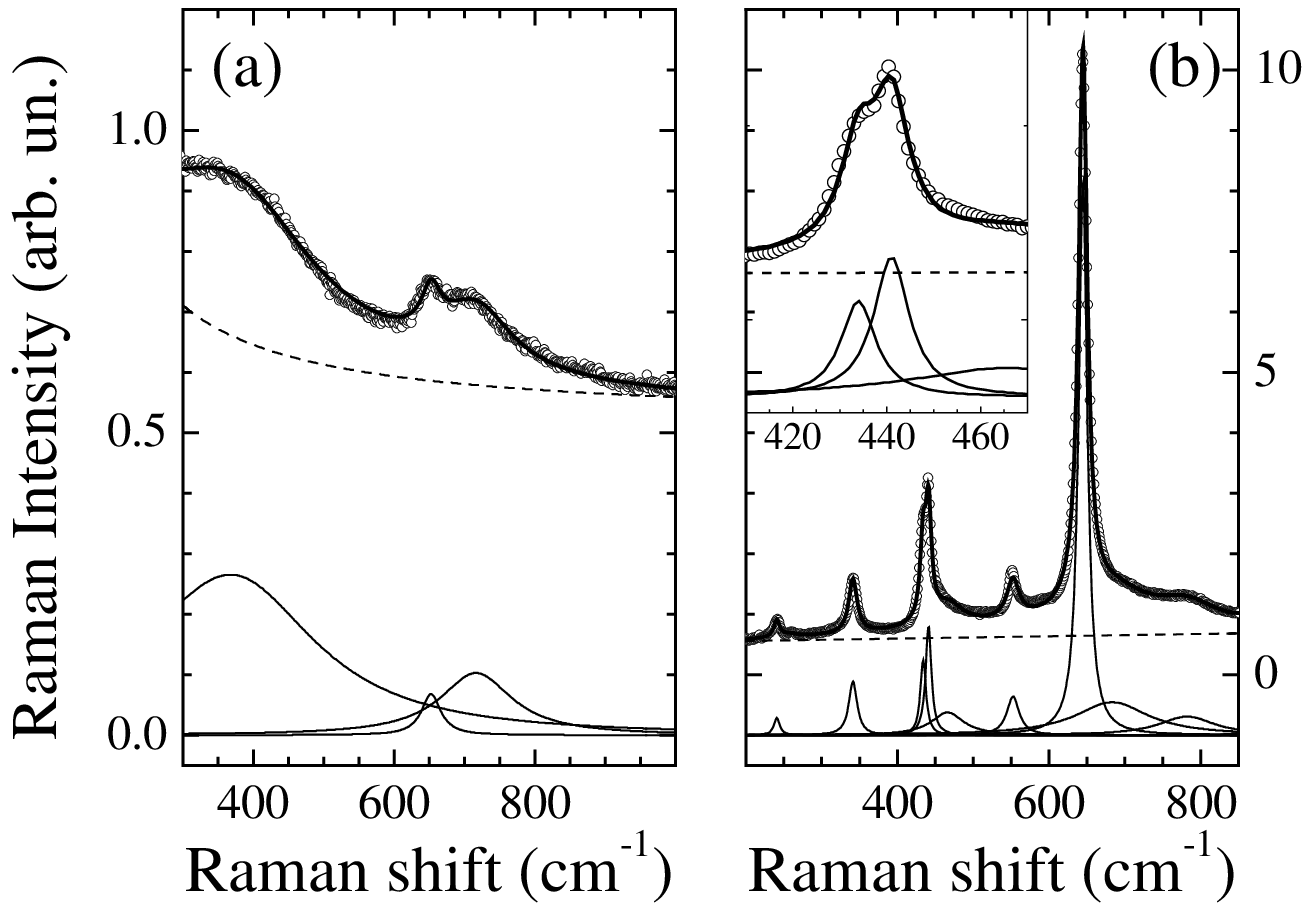}
\caption{Raman spectrum of cubic (a) and hexagonal (b) SMO (open
circles). Best-fit curves (thick solid lines), fitting components
(thin solid lines), and background contributions (dashed lines)
are also shown for both compounds. Fitting components in (b) have
been downshifted for sake of clarity. Note the different vertical
scale in (a) and (b). In (b) the inset shows the presence of a
double one-phonon peak at around 440 cm$^{-1}$.} \label{esacub}
\end{figure}

The typical Raman spectrum of the cSMO and a representative Raman
spectra of the hSMO are shown in Fig.~\ref{esacub}(a) and (b)
respectively. Bearing in mind that the intensities of the two
spectra are normalized to the same acquisition time, it is well
evident that the Raman signal of the cSMO sample is much less
structured and, on average, much smaller (roughly one order of
magnitude) than that of the hSMO (note the different vertical
scales in panels (a) and (b) of Fig.~\ref{esacub}).
This finding is consistent with group-theory which predicts $8$
Raman active phonons for the hexagonal symmetry ($P6_{3}/mmc$
space group) and none for the cubic perovskite structure. The weak
signal collected for the cubic sample is thus to be ascribed to
Raman-forbidden modes possibly activated by lattice disorder or
second-order Raman scattering. In the case of the hSMO compound,
which shows several well defined peaks in the Raman spectrum
(notice the two peaks structure in the inset of
Fig.~\ref{esacub}(b)), the total irreducible representation of the
Raman-active phonons can be determined starting from the
point-group symmetry, that is \cite{GroupTh}:
\begin{equation}
\Gamma_{Ram}=2A_{1g}+2E_{1g}+4E_{2g} \label{GammaRam}
\end{equation}
The high signal to noise ratio of the data allows for a careful
lineshape analysis. Both spectra were fitted by means of a linear
combination of damped harmonic oscillators (DHO) and a background
term \cite{diffusive,diffusive2}:
\begin{equation}
I(\nu)=I_{BKG}(\nu)+\sum_{i=1}^N {\frac{A_i \Gamma_i^2 \nu_i
\nu}{(\nu^2-\nu_i^2)^2+\Gamma_i^2 \nu^2}}
\end{equation}
where $I_{BKG}(\nu)$ is the background contribution and the second
term is the contribution from $N$ phonon peaks of frequency
$\nu_i$, width $\Gamma_i$, and amplitude $A_i$. We used $N=12$
phonon peaks plus a linear $I_{BKG}(\nu)$ for the cSMO and $N=3$
phonon peaks plus a diffusive electronic $I_{BKG}(\nu)$
\cite{diffusive,diffusive2} for the cubic compound. The best-fit
curves and single fitting-components are also shown in
Fig.~\ref{esacub}, whereas the corresponding best-fit phonon
frequencies $\nu_i$ and widths $\Gamma_i$, are reported in
tab.~\ref{fit}.

\begin{table}[b]
\centering
\begin{tabular}{|c|c|}
\hline
\multicolumn{2}{|c|}{\textbf{Cubic}}\\
\hline $\nu_i$ (cm$^{-1}$) & $\Gamma_i$ (cm$^{-1}$) \\
\hline 405 & 296 \\
652 & 33 \\ 719 & 125 \\ \hline
\end{tabular}
\begin{tabular}{|c|c|c|c|}
\hline
\multicolumn{4}{|c|}{\textbf{Hexagonal}}\\
\hline
$\nu_i$ (cm$^{-1}$) & $\Gamma_i$ (cm$^{-1}$) & 1-ph/m-ph & Label \\
\hline
243 & 9 & 1-ph & P$_1$ \\
344 & 13 & 1-ph & P$_2$ \\
434 & 9 & 1-ph & P$_3$ \\
441 & 9 & 1-ph & P$_4$ \\
467 & 52 & m-ph & \\
552 & 20 & 1-ph & P$_5$ \\
644 & 14 & 1-ph & P$_6$ \\
686 & 119 & m-ph & \\
783 & 89 & m-ph & \\
\hline
\end{tabular}
\caption{Phonon frequencies $\nu_i$ and widths $\Gamma_i$ for cSMO
(top table) and for hSMO (bottom table). For hSMO 1-phonon (1-ph)
or multi-phonon (m-ph) contributions are also indicated.}
\label{fit}
\end{table}

Before focusing on the analysis of the hexagonal compound it is
worth to notice that it has been necessary to introduce the
diffusive term (see the dashed line in Fig.~\ref{esacub}(a)) due
to the scattering of charge carriers, only for the cubic sample
which shows a higher conductivity than that of the hexagonal
compound \cite{Conduc}.

\begin{figure}[!t]
\includegraphics[width=8cm]{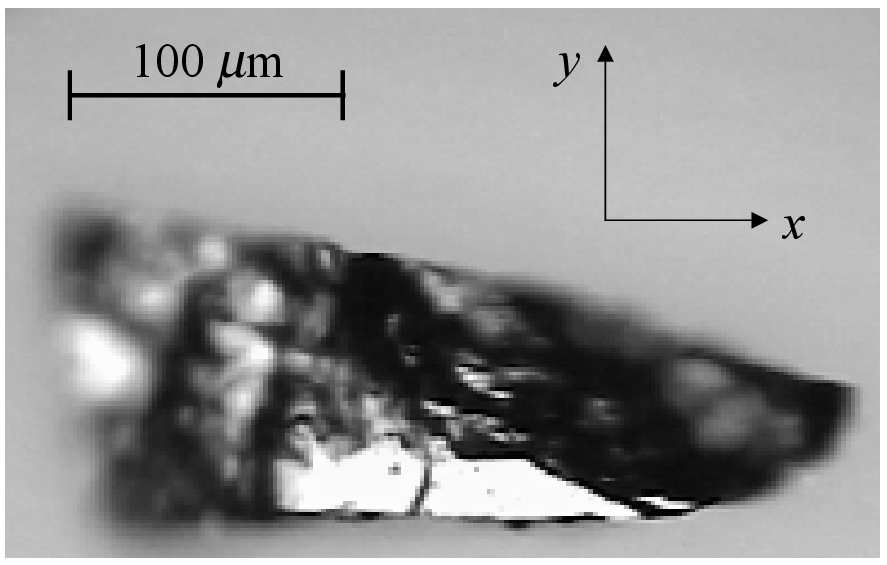}
\caption{Picture (taken with the microscope of the micro-Raman
apparatus) of the crystallite employed for polarization dependent
measurements.}\label{campione}
\end{figure}

\begin{figure}[!tb]
\includegraphics[width=8cm]{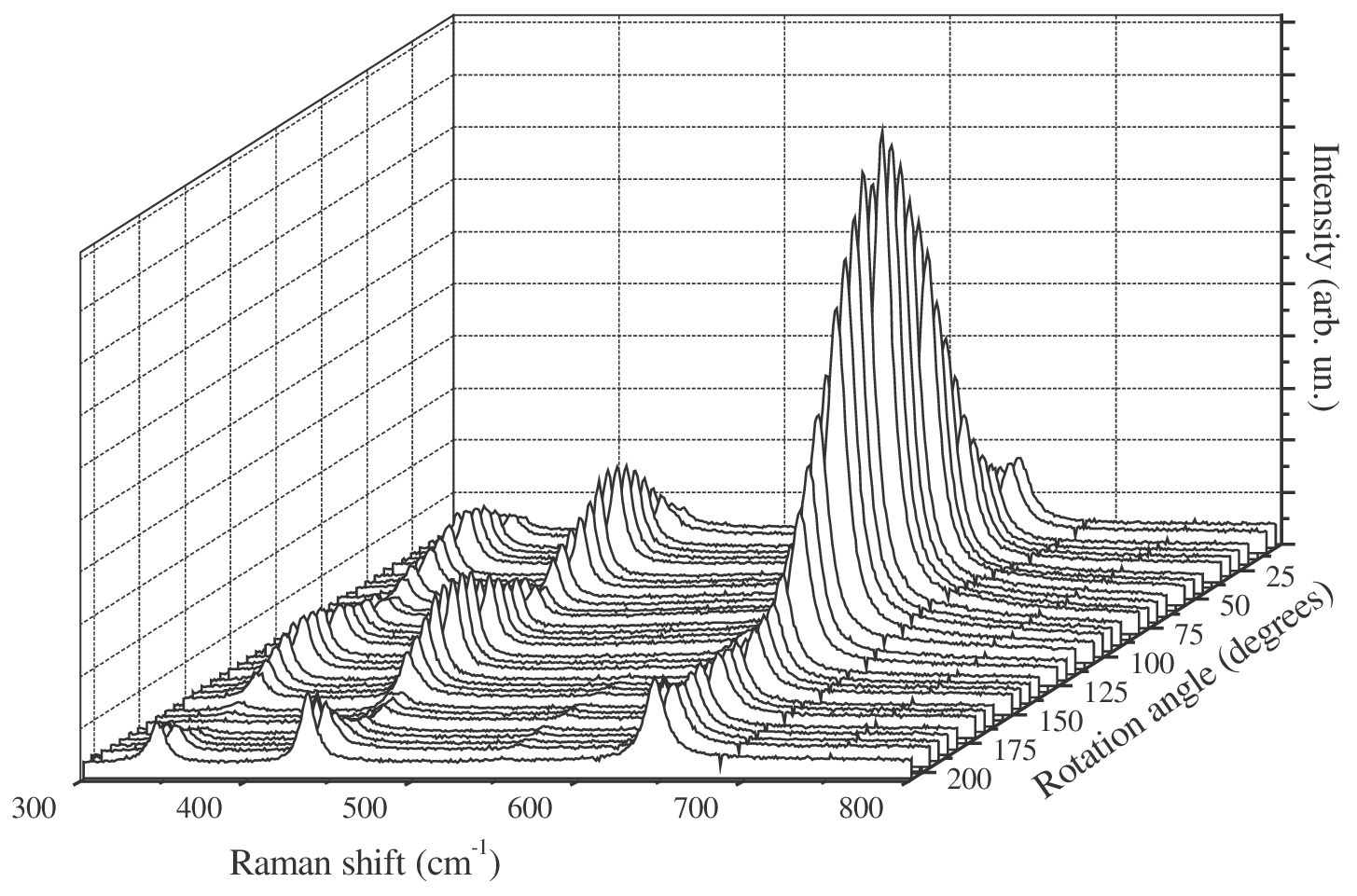}
\caption{Rotation-angle dependence of the Raman spectrum of the
crystallite shown in Fig.~\ref{campione}. The origin of the angle
scale is arbitrary.}\label{3D}
\end{figure}

Looking at the best-fit parameters reported in tab.~\ref{fit} for
the hexagonal sample we notice six rather narrow peaks ($\Gamma_i
\le 20$ cm$^{-1}$), labelled P$_1$-P$_6$ in increasing-frequency
order, and three broad peaks ($\Gamma_i > 50$ cm$^{-1}$). Since no
broadening factors are reasonably expected for the hSMO, we
tentatively ascribe the six P$_1$-P$_6$ peaks to one-phonon
processes out of the $8$ predicted, and the other ones to
multi-phonon processes. This assignment is confirmed and detailed
by the polarization analysis we are going to discuss. By
scratching the polycrystalline sample we were able to obtain
several crystallites. One of them, characterized by a flat surface
(see Fig.~\ref{campione}) which simple test measurements suggested
to be an $a-c$ plane, was chosen for the polarization analysis.
The flat surface of the crystallite was roughly aligned
perpendicular to the laser beam using the image from the
microscope. In order to carefully determine the orientation of the
crystallite, a preliminary set of measurements was performed,
rotating the crystallite from $0^\circ$ to $200^\circ$ with a
5$^\circ$ step. A configuration with scattered polarization
parallel to the incident one was selected. The angle dependence of
the spectrum is shown in Fig.~\ref{3D}. It can be noted that the
P$_{2}$ ($344$ cm$^{-1}$) and P$_{3}$ ($434$ cm$^{-1}$) peaks show
the same dependence and periodicity and thus the same symmetry,
whereas the P$_6$ ($644$ cm$^{-1}$) peak is characterized by a
clearly different behavior. The intensities (integrated areas) of
the P$_{2}$, P$_{3}$, and P$_{6}$ peaks were determined at each
angle by means of a single-peak fitting procedure. The resulting
angle-dependencies of the intensities of the three peaks were then
analyzed by means of three model functions based on group-theory
and determined as follows. We exploited the forms of the Raman
tensors for each allowed phonon symmetry \cite{GroupTh}:

\begin{equation}
A_{1g}: \left(
\begin{array}{ccc}
A & 0 & 0 \\
0 & A & 0 \\
0 & 0 & B \\
\end{array}
\right) \quad E_{1g}: \left(
\begin{array}{ccc}
0 & 0 & 0 \\
0 & 0 & C \\
0 & C & 0 \\
\end{array}\right),\left(
\begin{array}{ccc}
0 & 0 & C \\
0 & 0 & 0 \\
C & 0 & 0 \\
\end{array}\right)
\nonumber
\end{equation}
\begin{equation}
E_{2g}: \left(
\begin{array}{ccc}
-D & 0 & 0 \\
0 & D & 0 \\
0 & 0 & 0 \\
\end{array}\right),\left(
\begin{array}{ccc}
0 & D & 0 \\
D & 0 & 0 \\
0 & 0 & 0 \\
\end{array}\right)
\label{tensori}
\end{equation}

\noindent and the well-known relation for the Raman intensity:

\begin{equation}
    I= |\hat \varepsilon \alpha \hat \varepsilon'|^2
\end{equation}

\noindent where $\hat \varepsilon$ and $\hat \varepsilon'$ are the
polarization of the incident and the scattered light respectively
and $\alpha$ is the Raman tensor. Assuming an arbitrary tilting
angle $\beta$ between the laser beam and the crystal $c$-axis, for
each symmetry an analytical model function for the
angle-dependence of the intensity was determined. Since the
orientation of the crystal is unknown \emph{a priori} we have also
introduced an offset angle $\theta_{0}$ for the measured rotation
angle $\theta$ so that $\theta-\theta_0$ is the angle between the
incident polarization $\hat\varepsilon$ and the crystal $a$-axis.
The analytical expressions of the so-obtained model functions are
simple although long and they are not reported here.

Exploiting the calculated model functions, a fitting procedure was
applied to the angle dependence of the intensities of P$_2$,
P$_3$, and P$_6$ peaks, assuming $\beta$, $\theta_{0}$, and the
components of the considered Raman tensor assigned to each peak as
free fitting-parameters. Good quality fits were obtained only
assigning the $E_{1g}$ symmetry to the P$_{2}$ and P$_{3}$ peaks
and the $A_{1g}$ symmetry to the P$_{6}$ peak with a best fit
value of $72^\circ$ for $\beta$. This result confirms that the
flat surface of the crystallite is an $a-c$ plane and that
$\hat\varepsilon$ does not lie in this plane but forms a small
angle ($18^\circ$) with it. The $\theta_{0}$ best fit value
allowed us to conclude that the $c$ axis is parallel to the $y$
axis in Fig.~\ref{campione}.

Exploiting the orientation of the crystallite thus determined and
monitoring the Raman signal, we were able to carefully align the
flat $a-c$ surface of the crystallite, orthogonally to the laser
beam. In this case $\beta = 90^\circ$ and $\theta_{0}$ can be set
to $0^\circ$ so that the expressions for the angle-dependence of
the intensities of the phonon peaks are strongly simplified. Thus
in the parallel polarization geometry ($\hat \varepsilon =
(\cos\theta, 0, \sin\theta) = \hat \varepsilon'$ in the base of
crystal axes) one has:

\begin{eqnarray}\label{int}
I_{A_{1g}}^{\parallel} &=& (A-B)^{2}\cos^{4}\theta + 2B(A-B)\cos^{2}\theta + B^{2} \\
\nonumber I_{E_{1g}}^{\parallel}&=& 4C^{2}\sin^{2}\theta \cos^{2}\theta \\
\nonumber I_{E_{2g}}^{\parallel} &=& D^{2}\cos^{4}\theta
\end{eqnarray}

\noindent and in the orthogonal polarization geometry ($\hat
\varepsilon = (\cos\theta, 0, \sin\theta)$ and $\hat
\varepsilon'(-\sin\theta, 0, \cos\theta)$) one has:

\begin{eqnarray}\label{ort}
I_{A_{1g}}^{\perp} &=& (A-B)^{2} \sin^{2}\theta \cos^{2}\theta \\
\nonumber I_{E_{1g}}^{\perp} &=& C^{2} - 4C^{2}\sin^{2}\theta \cos^{2}\theta \\
\nonumber I_{E_{2g}}^{\perp} &=& D^{2}\sin^{2}\theta\cos^{2}\theta
\end{eqnarray}

\begin{figure}[!tb]
\includegraphics[width=8cm]{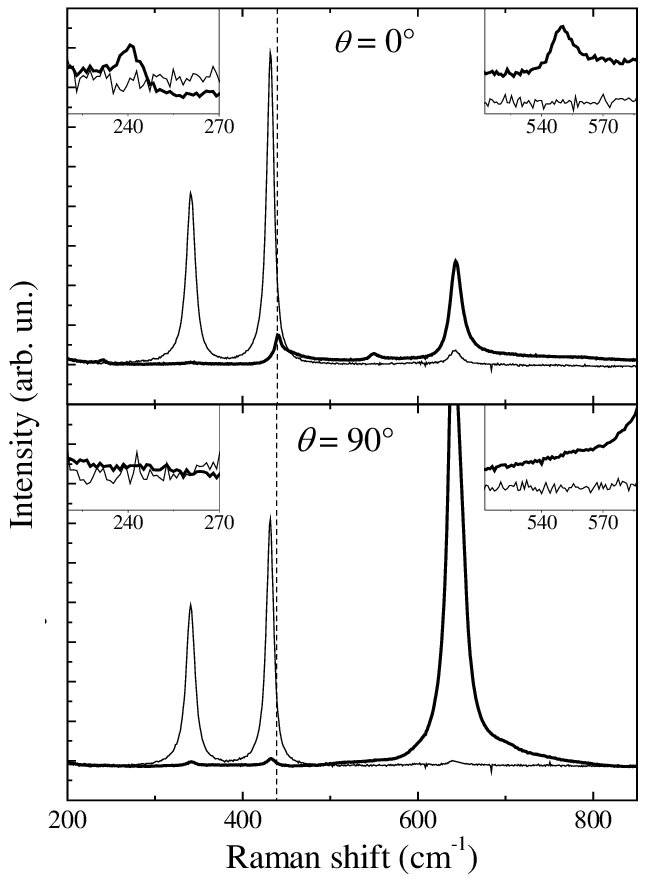}
\caption{Raman spectra of hSMO at $\theta = 0^\circ$ (top panel)
and $\theta= 90^\circ$ (bottom panel) in both parallel (thick
lines) and orthogonal (thin lines) configurations. Vertical dashed
lines mark the position of P$_4$ peak. Insets show magnifications
in the regions of P$_1$ and P$_5$ peaks.}\label{0}
\end{figure}

\begin{table}[!b]
\centering
\begin{tabular}{|c|c|c|}
\hline
& $\theta=0^\circ$ & $\theta=90^\circ$ \\
\hline
$I(A_{1g})^\|$ &$A^2$ & $B^2$ \\
$I(A_{1g})^\bot$ & 0 & 0 \\
$I(E_{1g})^\|$ & 0 & 0 \\
$I(E_{1g})^\bot$ & $C^2$ & $C^2$ \\
$I(E_{2g})^\|$ & $D^2$ & 0 \\
$I(E_{2g})^\bot$ & 0 & 0 \\
\hline
\end{tabular}
\caption{Calculated intensities of the three phonon symmetries in
both parallel ($\|$) and orthogonal ($\bot$) configurations at
high-symmetry angles. The constants $A$, $B$, $C$, and $D$ are the
components of the Raman tensors (see
eq.~\ref{tensori}).}\label{intensità}
\end{table}

\noindent with $A$, $B$, $C$ and $D$ elements of the Raman tensors
(see eq.~\ref{tensori}). We choose to analyze two high-symmetry
configurations, namely: $\theta=0^\circ$ ($\hat \varepsilon$
parallel to the $a$-axis) and $\theta=90^\circ$ ($\hat
\varepsilon$ parallel to the $c$-axis). The intensities for the
three symmetries $A_{1g}$, $E_{2g}$ and $E_{1g}$ can be calculated
in these high symmetry configurations exploiting eqs.~\ref{int}
and \ref{ort}, and the results are reported in
tab.~\ref{intensità}. Long acquisition-time spectra were then
collected for the selected configurations in order to obtain a
complete assignment of the observed phonon modes. Measurements,
shown in Fig.~\ref{0}, were performed in both parallel and
orthogonal polarization.
Comparing the data in Fig.~\ref{0} with tab.~\ref{intensità} the
assignment of the P$_{2}$, P$_{3}$, and P$_{6}$ peaks obtained
from the preliminary measurements is confirmed. Indeed, either for
$\theta=0^\circ$ or 90$^\circ$, P$_{2}$ and P$_{3}$ peaks are both
visible in orthogonal polarization and vanish in parallel
polarization as expected for the $E_{1g}$ symmetry, whereas
P$_{6}$ peak is apparent in the parallel polarization and vanishes
in the orthogonal polarization, as expected for the $A_{1g}$
symmetry. Four $E_{2g}$ modes and one $A_{1g}$ mode remain to be
assigned (see eq.~\ref{GammaRam}). The P$_{4}$ peak is clearly
visible in parallel polarization at $\theta=0^\circ$ and its
intensity drops to zero at $\theta = 90^\circ$, whereas in
orthogonal polarization it vanishes at both angles. Comparing this
behavior with tab.~\ref{intensità} we can assign the P$_{4}$ peak
to the $E_{2g}$ symmetry. The behavior of peaks P$_{1}$ and
P$_{5}$ is analogous (see the insets in Fig.~\ref{0}) to that of
the P$_{4}$ phonon and opposite to the P$_{6}$ peak. Therefore
P$_{1}$ and P$_{5}$ modes are also ascribed to the $E_{2g}$
symmetry. Thus the symmetry of all the six observed phonon peaks
was determined as reported in tab.~\ref{Assign}.

\begin{table}[!b]
\centering
\begin{tabular}{|c|c|c|c|}
\hline Label & $\nu_i$ (cm$^{-1}$) & Assignment & Atom \\
\hline P1 & 243 & $E_{2g}(2)$ & Mn \\
P2 & 344 & $E_{1g}(1)$ & Mn \\
P3 & 434 & $E_{1g}(2)$ & O \\
P4 & 441 & $E_{2g}(3)$ & O \\
P5 & 551 & $E_{2g}(4)$ & O \\
P6 & 644 & $A_{1g}(2)$ & O \\
\hline
\end{tabular}
\caption{Assignment of the 1-phonon peaks observed in hSMO.}
\label{Assign}
\end{table}

\begin{figure}[!tb]
\includegraphics[width=8cm]{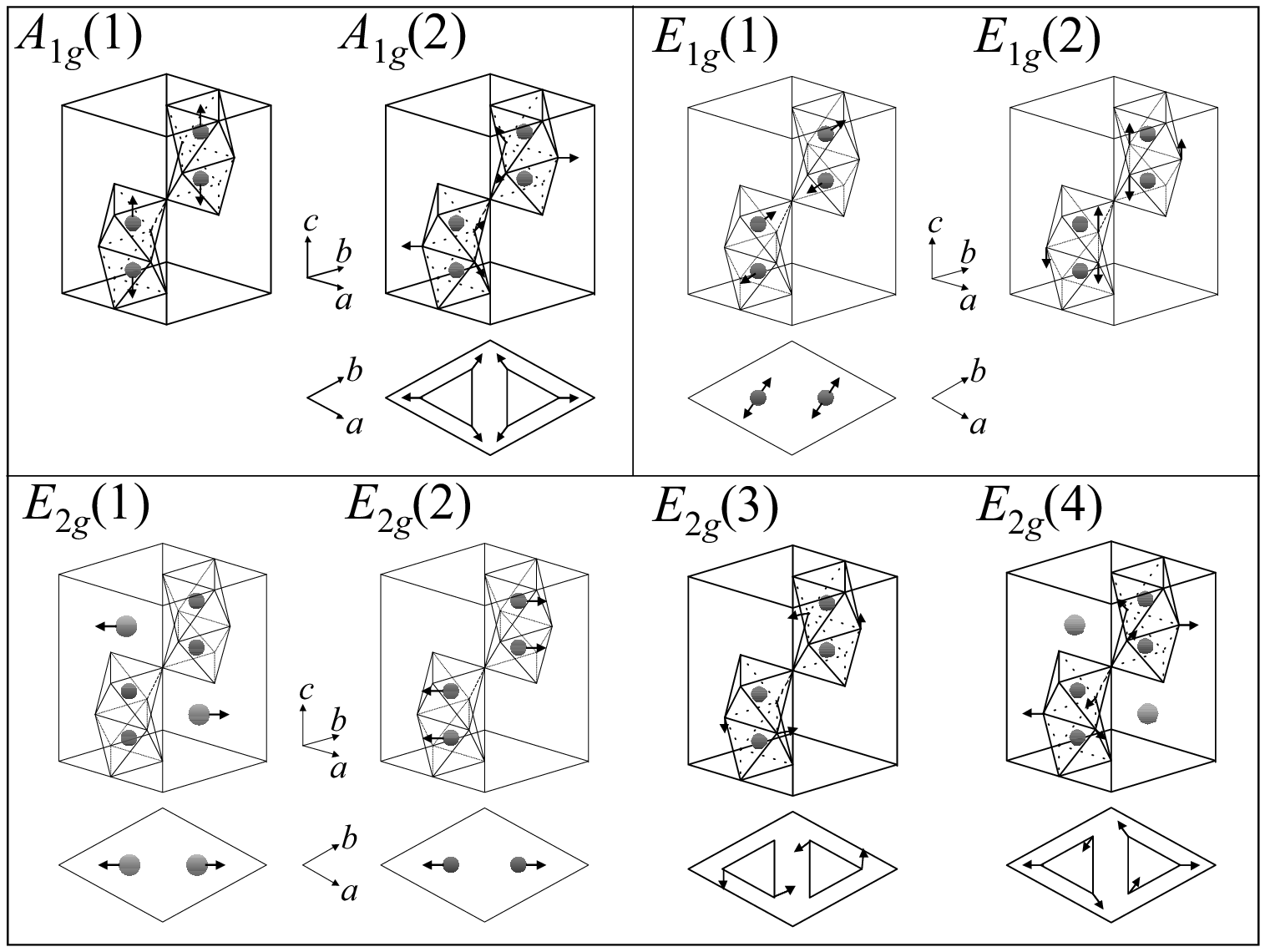}
\caption{Symmetry reduced atomic-displacements for Raman active
phonon modes in hSMO. Sr (Mn) atoms are indicated by large (small)
circles.}\label{modi}
\end{figure}

From the point-group symmetry of the crystal, the symmetry-reduced
coordinates for each phonon-symmetry can be determined (see
Fig.~\ref{modi}) \cite{GroupTh}.
Owing to the quite large difference between the atomic masses of
Sr, Mn, and O, it is reasonable to assume that the observed
phonons are almost pure modes, i.e. they mainly involve
displacements of a single atomic species. Therefore it is possible
to assign the phonon peaks appearing in the spectra to the atomic
displacements shown in Fig.~\ref{modi}. As a rule of thumb we can
assume that phonons involving displacements of lighter ions have
larger frequencies and moreover it is generally correct to assume
that stretching modes have frequencies higher than bending modes.
On the basis of these assumptions, we ascribe peaks P$_2$ and
P$_3$ to the $E_{1g}(1)$ (Mn displacements) and $E_{1g}(2)$ (O
displacements) modes respectively accordingly with the notation
used in Fig.~\ref{modi}. As to the P$_6$ peak ($A_{1g}$ symmetry)
its high vibrational frequency suggests the assignment of this
phonon to the $A_{1g}(2)$ mode (O displacements) rather than
$A_{1g}(1)$ mode (Mn displacements). This assignment is consistent
with that reported for isostructural BaRuO$_3$ \cite{BaRuO3}. The
remaining peaks P$_1$, P$_4$, and $P_5$ have $E_{2g}$ symmetry.
Bearing in mind that in BaRuO$_3$ the $E_{2g}(1)$ mode involving
Ba-displacements is centered at 90 cm$^{-1}$ (Ref.\cite{BaRuO3})
and that the ratio between atomic masses of Ba and Sr is 1.57, we
expect the $E_{2g}(1)$ phonon in hSMO to have a frequency well
below our experimental lower frequency-limit of 200 cm$^{-1}$.
Therefore, exploiting the aforementioned assumptions, we ascribe
P$_1$ peak to the $E_{2g}(2)$ mode (Mn displacement), whereas the
P$_4$ peak is assigned to the $E_{2g}(3)$ mode (O bending), and
the P$_5$ peak is ascribed to the $E_{2g}(4)$ mode (O stretching).

The assignment of the six one-phonon peaks observed in hSMO is
summarized in tab.~\ref{Assign}.

\section{CONCLUSIONS}

In summary, we reported the first optical characterization of both
the cubic and hexagonal SrMnO$_3$ manganite, performed by means of
Raman spectroscopy. Consistently with group-theory, which does not
predict Raman-active peaks in the cubic compound, the spectrum of
this compound substantially consists of very weak and broad
structures, probably due to Raman-forbidden modes and/or
multi-phonon processes. On the other hand, the spectrum of the
hexagonal compound shows six sharp peaks ascribed to one-phonon
processes. Exploiting our micro-Raman apparatus, a polarization
analysis was performed on a small crystallite of the hexagonal
sample. A preliminary set of measurements allowed us to assign the
symmetry of the main phonon peaks and to carefully determine the
orientation of the crystallite. A second set of measurements was
performed, aligning the crystallite with the $a-c$ plane
orthogonal to the optical axis. By using these data, a complete
assignment of all the observed Raman-active phonon peaks was
performed. This study by micro Raman on small manganite
crystallite show that a good accuracy can be achieved for these
transition metal oxides. It opens thus the route to the study of
other compounds such as YMnO$_3$ for instance in which the
trivalent manganese adopts a five fold coordination in MnO$_5$
trigonal bipyramides.


\begin{thebibliography}{99}

\bibitem{generali} Millis AJ. \emph{Nature.} 1998; \textbf{392}, 147.

\bibitem{generali2} Fontcuberta J. \emph{Phys. World.} 1999; \textbf{Feb}: 33.

\bibitem{dagotto} Dagotto E, Hotta T, Moreo A. \emph{Phys. Rep.} 2001;
\textbf{344}: 1.

\bibitem{raman} Postorino P, Congeduti A, Degiorgi E, Iti\'e JP, Munsch P.
\emph{Phys. Rev. B.} 2002; \textbf{65}: 224102. Dediu V,
Ferdeghini C, Matacotta FC, Nozar P, Ruani G. \emph{Phys. Rev.
Lett.} 2000; \textbf{84}: 4489. Iliev MN, Abrashev MV, Lee HG,
Popov VN, Sun YY, Thomsen C, Meng RL, Chu CW. \emph{Phys. Rev. B.}
1998; \textbf{57}: 2872.

\bibitem{Hervieu} Hervieu M, Martin C, Maignan A, Van Tendeloo G, Jirak Z,
Hejtmanek J, Barnab\'e A, Thopart D, Raveau B. \emph{Chem. Mater.}
2000; \textbf{12}: 1456.

\bibitem{Chmaissem} Chmaissem O, Dabrowski B, Kolesnik S, Mais J,
Brown DE, Kruk R, Prior P, Pyles B, Jorgensen JD. \emph{Phys Rev
B.} 2001; \textbf{64}: 134412.


\bibitem{Battle} Battle PD, Gibb TC, Jones CW. \emph{J. Solid State
Chem.} 1998; \textbf{74}: 60.

\bibitem{takeda} Takeda T, Ohara S. \emph{J. Phys. Soc. Jpn.}
1974; \textbf{37}: 275.


\bibitem{GroupTh} Fateley WG, Dollish FR, McDevitt HT, Bentley FF.
\emph{Infrared and Raman Selection Rules for Molecular and Lattice
Vibrations: The Correlation Method}; Wiley-Interscience: New York,
USA, 1972.

\bibitem{diffusive} Yoon S, Liu HL, Schollerer G, Cooper SL, Han PD, Payne DA, Cheong SW,
Fisk Z. \emph{Phys. Rev. B.} 1998; \textbf{58}: 2795.

\bibitem{diffusive2} Congeduti A, Postorino P, Caramagno E, Nardone M, Kumar A,
Sarma DD. \emph{Phys. Rev. Lett.}; 2001; \textbf{86}: 1251.

\bibitem{Conduc} Hashimoto S, Iwahara H. \emph{J. Electroceram.} 2000;
\textbf{1}: 225.

\bibitem{BaRuO3} Lee YS, Noh TW, Park JH, Lee KB, Cao G, Crow JE, Lee MK, Eom CB, Oh EJ, In-Sang
Yang. \emph{Phys. Rev. B} 2002; \textbf{65}: 235113.

\end{thebibliography}
\end{document}